\documentclass[twocolumn,preprintnumbers,amsmath,amssymb]{revtex4}

\usepackage{bm}
\usepackage{dcolumn}
\usepackage{epsfig}
\usepackage{graphicx}

\begin{document}


\title{Possibilities for reduction of transverse projected emittances\\
by partial removal of transverse to longitudinal beam correlations}

\author{V.Balandin}
\email{vladimir.balandin@desy.de}
\author{W.Decking}
\email{winfried.decking@desy.de}
\author{N.Golubeva}
\email{nina.golubeva@desy.de}
\affiliation{
Deutsches Elektronen-Synchrotron DESY, Notkestrasse 85, 22607 Hamburg, Germany
}%

\date{July 24, 2014}

\begin{abstract}
We show that if in the particle beam there are linear correlations between energy 
of particles and their transverse positions and momenta (linear beam dispersions), 
then the transverse projected emittances always can be reduced by letting the beam 
to pass through magnetostatic system with specially chosen nonzero lattice dispersions.
The maximum possible reduction of the transverse projected emittances occurs when all 
beam dispersions are zeroed, and the values of the lattice dispersions required for that
are completely defined by the values of the beam dispersions and the beam rms energy 
spread and are independent from any other second-order central beam moments.
Besides that, we prove that, alternatively, one can also use the lattice dispersions to
remove linear correlations between longitudinal positions of particles and their 
transverse coordinates (linear beam tilts), but in this situation solution for the 
lattice dispersions is nonunique and the reduction of the transverse projected 
emittances is not guaranteed.
\end{abstract}


\maketitle

\section{Introduction}

Careful control of the beam quality is essential for linear accelerators designed 
to deliver very high brightness electron beams for short wavelength free electron 
lasers (FELs). There are many beam properties which have to be observed and manipulated
such as suppression of microbunching instability, creation of needed peak current, 
preservation of slice and projected emittances, etc. 

In this paper we are interested in some aspects of the control of transverse projected 
emittances. Among the sources of the growth of transverse projected emittances are 
the incoherent and coherent synchrotron radiation (CSR) withing magnetic bunch 
compressors as well as the other wake fields along the accelerator. A number of 
approaches which could help to reduce emittance growth due to CSR wake during bunch 
compression were developed during last decades including different optics tricks, 
preparation of the initial beam current profile at the bunch compressor entrance 
and etc. (see, for example, ~\cite{LouMos,DohL,SdiM,JHL,MQE} and references therein). 

Still, because the suggested schemes provide reduction but not complete cancellation 
of the emittance growth, the beam considered downstream of the compression system 
(or at the linac exit) could have nonzero transverse to longitudinal coupling terms 
in the beam matrix and therefore projected emittances could be further reduced if
these correlations will be removed.

In general, in order to make complete transverse to longitudinal decoupling, it is 
necessary to have the possibility to act on particles depending on their longitudinal 
position within the bunch (for example, one may involve transverse deflecting cavities
for this purpose), which means that the system designed for the complete decoupling 
could be too complicated and somewhat difficult to operate in comparison with the
benefit coming from the achievable reduction of the transverse projected emittances. 

In this paper we consider a more simple and more practical question: what one can do 
having at hand a magnetostatic correction system? Because the transfer matrix of 
a magnetostatic system could couple transverse and longitudinal particle coordinates 
only when the dispersions of the underlying magnetic structure are nonzero, the reduction 
of transverse projected emittances (if any possible) will always be accompanied by 
the creation of a potential source of beam transverse jitter due to the beam 
energy jitter, and one has to look for an appropriate balance of both.

We show, in the framework of linear particle dynamics and with the self field effects 
neglected, that if in the beam matrix there are nonzero correlation terms between energy 
of particles and their transverse positions and momenta (beam dispersions), then the 
transverse projected emittances can be reduced by letting the beam pass through 
magnetostatic system (correction system) with specially chosen nonzero lattice dispersions.
The maximum possible reduction of the transverse projected emittances occurs when all 
beam dispersions are zeroed, and the values of the lattice dispersions required for that
are completely determined by the values of the beam dispersions and the beam rms energy 
spread and are independent from any other second-order central beam moments.
Besides that, we prove that, alternatively, one can also use the lattice dispersions to
remove linear correlations between longitudinal positions of particles and their 
transverse coordinates (beam tilts), but in this situation solution for the lattice 
dispersions is nonunique and the reduction of the transverse projected emittances is not 
guaranteed.
 
Note that this paper is an extended version of the unpublished note \cite{My00},
which was written during discussion of the influence of different dispersive effects
on the performance of the FLASH facility \cite{Flash01,Flash02}, and recently,
when we get acquainted with the paper \cite{PSI}, our interest to this problem was renewed.
Both papers, \cite{My00} and \cite{PSI}, employ unclosed lattice dispersions as a tuning 
knob for the control of the transverse projected emittances, but have somewhat different 
points of view on the practical realization of this idea and therefore the direct comparison 
of their results and recommendations is not very straightforward. As far as we are mostly 
discussing possibilities for correction which can be made downstream of the emittance growth 
and coupling source either by means of a dedicated correction system or even simply by
special (dispersive) beam steering, the paper \cite{PSI} suggests to create dispersion 
nonclosure already in the bunch compressor, where the CSR effect is strongest and can 
not be neglected.

\section{Variables and notations}

We consider the linear beam dynamics in an electromagnetic system which conserves 
the reference beam energy and take the path length along the reference orbit $\tau$ 
to be the independent variable. We use a complete set of symplectic variables 

\noindent
\begin{eqnarray}
\mbox{\boldmath $z$} = (x, p_x, y, p_y, \sigma, \varepsilon)^{\top}
\label{Intr01} 
\end{eqnarray}

\noindent
as particle coordinates \cite{MaisRipken, My01}. Here $x, \, y$ measure 
the transverse (horizontal and vertical) displacements from the ideal 
orbit and $p_x, \, p_y$ are the corresponding canonical monenta scaled 
with the constant kinetic momentum  of the reference particle $p_0$. 
The variables $\sigma$ and $\varepsilon$ which describe the longitudinal 
dynamics are 

\noindent
\begin{eqnarray}
\sigma =  c \,\beta_0 \, (t_0 - t),
\quad
\varepsilon =
({\cal{E}} - {\cal{E}}_0) \, / \, (\beta_0^2 \,{\cal{E}}_0),
\label{L10_0} 
\end{eqnarray}

\noindent
where ${\cal{E}}_0, \, \beta_0$ and $t_0 = t_0(\tau)$ are the energy of the 
reference particle, its velocity in terms of the speed of light $c$ and its 
arrival time at a certain position $\tau$, respectively.

Let $M$ be an $m \times m$ square matrix. Then $\left| M \right|$ denote 
the determinant of $M$. Let $\omega$ be a nonempty subset of $\{1,2,\ldots,m\}$ 
with its elements listed in increasing order. Then $M\{\omega\}$ denote 
the principal submatrix of $M$ whose entries are in the intersection of 
those rows and columns of $M$ specified by $\omega$. If $M$ is a symmetric matrix, 
we denote by $\Psi_M$ the associated with this matrix quadratic form
in $m$-variables $u_1, \ldots, u_m$

\noindent
\begin{eqnarray}
\Psi_M (u_1, \ldots, u_m) =
(u_1, \ldots, u_m) \cdot M \cdot (u_1, \ldots, u_m)^{\top}.
\label{quadrFormA}
\end{eqnarray}

\noindent
Besides that, we denote by $I_m$ the $m \times m$ identity matrix and by

\noindent
\begin{eqnarray}
J_{2m} \;=\;\mbox{diag}
\Bigg(
\underbrace{
\left(
\begin{array}{rr}
0 & 1\\
-1 & 0
\end{array}
\right),
\ldots,
\left(
\begin{array}{rr}
0 & 1\\
-1 & 0
\end{array}
\right)
}_{m}
\Bigg)
\label{simplUnit}
\end{eqnarray}

\noindent
the $2m \times 2m$ symplectic unit matrix.

As usual, we describe the properties of a collection of points (a particle beam)
in the three degrees of freedom (3D) phase space by a $6 \times 6$ symmetric 
matrix (beam matrix) of the second-order central beam moments

\noindent
\begin{eqnarray}
\Sigma =
\left\langle
\left(\mbox{\boldmath $z$} - \langle \mbox{\boldmath $z$} \rangle \right) 
\left(\mbox{\boldmath $z$} - \langle \mbox{\boldmath $z$} \rangle \right)^{\top}
\right\rangle,
\label{intr_2}
\end{eqnarray}

\noindent
where the brackets $\langle \, \cdot \, \rangle$ denote an average over 
a distribution of the particles in the beam. 

Let $R$ be the nondegenerated $6 \times 6$ matrix which propagates particle 
coordinates from the state $\tau = s_1$ to the state $\tau = s_2$, i.e let

\noindent
\begin{eqnarray}
\mbox{\boldmath $z$}(s_2) \, = \, R\, \mbox{\boldmath $z$}(s_1).
\label{intr_4}
\end{eqnarray}

\noindent
Then from (\ref{intr_2}) and (\ref{intr_4}) it follows that the matrix $\Sigma$ 
evolves between these two states according to the congruence

\noindent
\begin{eqnarray}
\Sigma(s_2) \,=\, R \, \Sigma(s_1) \, R^{\top}.
\label{Form07}
\end{eqnarray}

\noindent
In the following we assume that the beam transport matrix $R$ is symplectic, 
which is equivalent to say that it satisfies the relation

\noindent
\begin{eqnarray}
R^{\top} J_6 \, R \, = \, J_6.
\label{Form07SymplCond}
\end{eqnarray}

By definition, the beam matrix $\Sigma$ is symmetric positive semidefinite and 
we restrict our considerations to the situation when this matrix is nondegenerated 
and therefore positive definite. For simplification of notations we also assume 
that the beam is proper centered and therefore has vanishing first-order 
moments $\big< \mbox{\boldmath $z$} \big> = 0$. With this assumption the beam 
matrix takes on the form

\noindent
\begin{eqnarray}
\Sigma = 
\left(
\begin{array}{cccccc}
\langle x^2                \rangle & \langle x p_x              \rangle & 
\langle x y                \rangle & \langle x p_y              \rangle & 
\langle x \sigma           \rangle & \langle x \varepsilon      \rangle\\
\langle x p_x              \rangle & \langle p_x^2              \rangle & 
\langle y p_x              \rangle & \langle p_x p_y            \rangle & 
\langle p_x \sigma         \rangle & \langle p_x \varepsilon    \rangle\\
\langle x y                \rangle & \langle y p_x              \rangle & 
\langle y^2                \rangle & \langle y p_y              \rangle & 
\langle y \sigma           \rangle & \langle y \varepsilon      \rangle\\
\langle x p_y              \rangle & \langle p_x p_y            \rangle & 
\langle y p_y              \rangle & \langle p_y^2              \rangle & 
\langle p_y \sigma         \rangle & \langle p_y \varepsilon    \rangle\\
\langle x \sigma           \rangle & \langle p_x \sigma         \rangle & 
\langle y \sigma           \rangle & \langle p_y \sigma         \rangle & 
\langle \sigma^2           \rangle & \langle \sigma \varepsilon \rangle\\
\langle x \varepsilon      \rangle & \langle p_x \varepsilon    \rangle & 
\langle y \varepsilon      \rangle & \langle p_y \varepsilon    \rangle & 
\langle \sigma \varepsilon \rangle & \langle \varepsilon^2      \rangle
\end{array}
\right)
\label{Form06}
\end{eqnarray}

\noindent
where the elements 

\noindent
\begin{eqnarray}
\langle x   \varepsilon \rangle, 
\quad
\langle p_x \varepsilon \rangle,
\quad 
\langle y   \varepsilon \rangle, 
\quad
\langle p_y \varepsilon \rangle
\label{Form067}
\end{eqnarray}

\noindent
and the elements 

\noindent
\begin{eqnarray}
\langle x   \sigma \rangle, 
\quad
\langle p_x \sigma \rangle,
\quad 
\langle y   \sigma \rangle, 
\quad
\langle p_y \sigma \rangle 
\label{Form068}
\end{eqnarray}

\noindent
we call beam dispersions and beam tilts, respectively.

The matrix $\Sigma$ has twenty-one different entries which can be varied independently 
within the positive definiteness conditions. Of course, not all of them (or their 
combinations) are equally interesting for any particular accelerator physics application.
In this paper we concentrate on the study of the evolution of 1D horizontal, 
vertical and longitudinal projected emittances

\noindent
\begin{eqnarray}
\varepsilon_x = \left| \Sigma \left\{ 1, 2 \right\} \right|^{1 / 2},
\label{E1Dex}
\end{eqnarray}

\noindent
\begin{eqnarray}
\varepsilon_y = \left|\Sigma\left\{3,4\right\} \right|^{1 / 2},
\label{E1Dey}
\end{eqnarray}

\noindent
\begin{eqnarray}
\varepsilon_{\sigma} = \left|\Sigma\left\{5,6\right\} \right|^{1 / 2},
\label{E1Des}
\end{eqnarray}

\noindent
and 2D transverse projected emittance

\noindent
\begin{eqnarray}
\varepsilon_t = \left|\Sigma\left\{1,2,3,4\right\} \right|^{1 / 2}
\label{E1Det}
\end{eqnarray}

\noindent
under the transformation rule (\ref{Form07}) with the additional assumption 
that the matrix $R$ is the transport matrix of a magnetostatic system. Besides 
that, we also pay attention to the changes in the beam energy chirp 
$\langle \sigma \varepsilon \rangle$ (linear energy slope along the bunch length)
and in the rms bunch length squared $\langle \sigma^2 \rangle$.

Note that due to Hadamar's determinantal inequality

\noindent
\begin{eqnarray}
\varepsilon_t \, \leq \, \varepsilon_x \, \varepsilon_y,
\label{EInDet}
\end{eqnarray}

\noindent
and the equality in (\ref{EInDet}) holds if and only if
the transverse degrees of freedom in the beam matrix are decoupled 
from each other \cite{HornJohnson}, i.e. if and only if 

\noindent
\begin{eqnarray}
\langle x y \rangle
=
\langle x p_y \rangle
=
\langle y p_x \rangle
=
\langle p_x p_y \rangle
=
0.
\label{noCouplXY678}
\end{eqnarray}

\section{Transport of beam matrix through magnetostatic system}

\subsection{Matrix of a magnetostatic system}

The most general 
form of the transport matrix of a magnetostatic system is

\noindent
\begin{eqnarray}
R = 
\left(
\begin{array}{cccccc}
r_{11} & r_{12} & r_{13} & r_{14} & 0 & r_{16}\\
r_{21} & r_{22} & r_{23} & r_{24} & 0 & r_{26}\\
r_{31} & r_{32} & r_{33} & r_{34} & 0 & r_{36}\\
r_{41} & r_{42} & r_{43} & r_{44} & 0 & r_{46}\\
r_{51} & r_{52} & r_{53} & r_{54} & 1 & r_{56}\\
0      & 0      & 0      & 0      & 0 & 1
\end{array}
\right),
\label{Form01}
\end{eqnarray}

\noindent
where the elements 

\noindent
\begin{eqnarray}
r_{16}, \; 
r_{26}, \; 
r_{36}, \; 
r_{46}, \; 
r_{51}, \; 
r_{52}, \; 
r_{53}, \; 
r_{54}, \; 
r_{56}
\label{Form01LattDisp}
\end{eqnarray}

\noindent
are (transverse and longitudinal) lattice dispersions. 

The special form (\ref{Form01}) of the matrix $R$ allows to rewrite 
the symplecticity condition (\ref{Form07SymplCond}) in the form of a 
system of two equations

\noindent
\begin{eqnarray}
\left(R\{1,2,3,4\}\right)^{\top}
J_4\,
\left(R\{1,2,3,4\}\right) \, = \, J_4
\label{Form02}
\end{eqnarray}

\noindent
and

\noindent
\begin{eqnarray}
\left(
\begin{array}{c}
r_{16}\\
r_{26}\\
r_{36}\\
r_{46}
\end{array}
\right)
\, = \,
\left(R\{1,2,3,4\}\right)
J_4
\left(
\begin{array}{c}
r_{51}\\
r_{52}\\
r_{53}\\
r_{54}
\end{array}
\right),
\label{Form022}
\end{eqnarray}

\noindent
and using condition (\ref{Form022}) one can show 
that every matrix $R$ of the form (\ref{Form01}) can be represented 
as a product  

\noindent
\begin{eqnarray}
R \, = \, R_1 \, R_2,
\label{Form03}
\end{eqnarray}

\noindent
where

\noindent
\begin{eqnarray}
R_1 = 
\left(
\begin{array}{cccccc}
r_{11} & r_{12} & r_{13} & r_{14} & 0 & 0\\
r_{21} & r_{22} & r_{23} & r_{24} & 0 & 0\\
r_{31} & r_{32} & r_{33} & r_{34} & 0 & 0\\
r_{41} & r_{42} & r_{43} & r_{44} & 0 & 0\\
0      & 0      & 0      & 0      & 1 & 0\\
0      & 0      & 0      & 0      & 0 & 1
\end{array}
\right)
\label{Form04}
\end{eqnarray}

\noindent
is the dispersion-free part of the matrix $R$ and

\noindent
\begin{eqnarray}
R_2 = 
\left(
\begin{array}{cccccc}
1      & 0      & 0      & 0      & 0 & \;\;\,r_{52}\\
0      & 1      & 0      & 0      & 0 & -r_{51}\\
0      & 0      & 1      & 0      & 0 & \;\;\,r_{54}\\
0      & 0      & 0      & 1      & 0 & -r_{53}\\
r_{51} & r_{52} & r_{53} & r_{54} & 1 & \;\;\,r_{56}\\
0      & 0      & 0      & 0      & 0 & \;\;\,1
\end{array}
\right)
\label{Form05}
\end{eqnarray}

\noindent
is its dispersive part.

Substituting the decomposition (\ref{Form03}) into
the beam matrix propagation equation (\ref{Form07}) one obtains

\noindent
\begin{eqnarray}
\Sigma(s_2) 
\,=\, 
R_1 \, \left( R_2 \, \Sigma(s_1) \, R_2^{\top} \right) R_1^{\top}.
\label{Form079997}
\end{eqnarray}

\noindent
This formula is a two step transformation. At first the incoming beam 
matrix $\Sigma(s_1)$ is transported using the matrix $R_2$ and then this 
intermediate result is transformed using the matrix $R_1$. Because the action 
of the matrix $R_1$ does not alter longitudinal beam parameters, does not 
couple transverse and longitudinal projected emittances, and propagates the vector 
of beam dispersions and the vector of beam tilts simply as transverse coordinates 
of the particle trajectories  (i.e. without possibilities to create or to remove 
vectors of beam dispersions and beam tilts, and even without possibility simply 
to mix the vector of the beam dispersions with the vector of the beam tilts), 
the second step in the transport of the beam matrix can be omitted without 
loss of generality for any result of this paper. So, in the rest of this paper, 
we consider the changes in properties of the incoming 
beam matrix $\Sigma(s_1)$ which are of interest for us
under the action of the matrix $R_2$ only. Because 
it is impossible to associate with this action some certain position in 
the beam line, we write it symbolically as follows

\noindent
\begin{eqnarray}
\Sigma \, \leftarrow \,
R_2 \, \Sigma \, R_2^{\top},
\label{S01}
\end{eqnarray}

\noindent
and call this transformation as the beam passage through 
the dispersive part of the correction system.

Note that the formulas obtained below 
for the simplified propagation rule (\ref{S01}) can be translated 
into the formulas for the complete transport equation (\ref{Form07}) with 
the help of the decomposition of the matrix $R$ in the form of a product

\noindent
\begin{eqnarray}
R \,=\, R_3 \,R_1,
\label{Form033}
\end{eqnarray}

\noindent
where

\noindent
\begin{eqnarray}
R_3 = 
\left(
\begin{array}{cccccc}
\;\;\,1 & 0      & \;\;\,0      & 0      & 0 & r_{16}\\
\;\;\,0 & 1      & \;\;\,0      & 0      & 0 & r_{26}\\
\;\;\,0 & 0      & \;\;\,1      & 0      & 0 & r_{36}\\
\;\;\,0 & 0      & \;\;\,0      & 1      & 0 & r_{46}\\
-r_{26} & r_{16} & -r_{46}      & r_{36} & 1 & r_{56}\\
\;\;\,0 & 0      & \;\;\,0      & 0      & 0 & 1
\end{array}
\right),
\label{Form053}
\end{eqnarray}

\noindent
and the matrix $R_1$ remains the same as given in (\ref{Form04}).
To make such a translation in the selected formula one has to make 
the following changes in its right hand side:
substitute the lattice dispersions $r_{16}$, $r_{26}$, $r_{36}$, and $r_{46}$
instead of the lattice dispersions $r_{51}$, $r_{52}$, $r_{53}$, and $r_{54}$
according to the rule

\noindent
\begin{eqnarray}
r_{51} \rightarrow -r_{26},
\;
r_{52} \rightarrow r_{16},
\;
r_{53} \rightarrow -r_{46},
\;
r_{54} \rightarrow r_{36},
\label{Form064253}
\end{eqnarray}

\noindent
and substitute the elements of the matrix $R_1 \Sigma(s_1) R_1^{\top}$
instead of the corresponding elements of the matrix $\Sigma(s_1)$.

Note that the decompositions (\ref{Form03}) and (\ref{Form033}) are 
still valid if one simply shifts the element $r_{56}$ from 
the matrices $R_2$ and $R_3$ to the corresponding position in the matrix $R_1$.
It gives an additional possibility to simplify calculations if one cares about
transport of the projected emittances only, but
because we are also interested in the behavior of the rms bunch length
and the beam energy chirp, we prefer to keep the $r_{56}$ coefficient in 
the matrices $R_2$ and $R_3$.

\subsection{Transformation of 1D projected emittances}

In order to obtain convenient representation for the emittance 
transport problem, let us introduce a $4 \times 4$ symmetric matrix

\noindent
\begin{eqnarray}
A 
=
\langle \varepsilon^2 \rangle
\Sigma\left\{1,2,3,4\right\}
-
\left(
\begin{array}{c}
\langle x \varepsilon \rangle\\
\langle p_x \varepsilon \rangle\\
\langle y \varepsilon \rangle\\
\langle p_y \varepsilon \rangle
\end{array}
\right)
\left(
\begin{array}{c}
\langle x \varepsilon \rangle\\
\langle p_x \varepsilon \rangle\\
\langle y \varepsilon \rangle\\
\langle p_y \varepsilon \rangle
\end{array}
\right)^{\top}.
\label{matrixA}
\end{eqnarray}

\noindent
Because leading principal minors of this matrix 
can be expressed through the principal minors of the
positive definite matrix $\Sigma$ as follows

\noindent
\begin{eqnarray}
\left| A\left\{1\right\} \right| =
\left| \Sigma\left\{1, 6\right\}\right|,
\label{Minor1}
\end{eqnarray}

\noindent
\begin{eqnarray}
\left| A\left\{1, 2\right\} \right| =
\left| \Sigma\left\{6\right\}\right| 
\left| \Sigma\left\{1, 2, 6\right\}\right|,
\label{Minor2}
\end{eqnarray}

\noindent
\begin{eqnarray}
\left| A\left\{1, 2, 3\right\} \right| =
\left| \Sigma\left\{6\right\}\right|^2 
\left| \Sigma\left\{1, 2, 3, 6\right\}\right|,
\label{Minor3}
\end{eqnarray}

\noindent
\begin{eqnarray}
\left| A\left\{1, 2, 3, 4\right\} \right| = 
\left| \Sigma\left\{6\right\}\right|^3 
\left| \Sigma\left\{1, 2, 3, 4, 6\right\}\right|,
\label{Minor4}
\end{eqnarray}

\noindent
all leading principal minors of the matrix $A$ are positive, which means that
the matrix $A$ is positive definite according to the Sylvester 
criterion \cite{HornJohnson}. 
Note that the elements of this matrix (similar 
to the elements of the matrix $B$ given below in (\ref{matrixB})) 
do not depend on the second-order beam moments involving the longitudinal 
variable $\sigma$.

With the help of the positive definite quadratic form $\Psi_A$
associated with the matrix $A$, the evolution of the 1D
projected emittances through the dispersive part of the correction system
can be expressed as follows:

\noindent
\begin{eqnarray}
\varepsilon_x^2 
\,\leftarrow\, 
\varepsilon_x^2
\,+\,
\Psi_A(r_{51}^x - r_{51}, \, r_{52}^x - r_{52}, \, 0, \, 0)
\nonumber
\end{eqnarray}

\noindent
\begin{eqnarray}
-\,
\Psi_A(r_{51}^x, \, r_{52}^x, \, 0, \, 0),
\label{evolutionEX}
\end{eqnarray}

\noindent
where

\noindent
\begin{eqnarray}
r_{51}^x 
\,=\, 
\frac{\langle p_x \varepsilon  \rangle}{\langle \varepsilon^2 \rangle},
\quad
r_{52}^x 
\,=\, 
-\frac{\langle x \varepsilon  \rangle}{\langle \varepsilon^2 \rangle}.
\label{optSolutionXX}
\end{eqnarray}

\noindent
\begin{eqnarray}
\varepsilon_y^2 
\,\leftarrow\, 
\varepsilon_y^2
\,+\,
\Psi_A(0, \, 0, \, r_{53}^y - r_{53}, \, r_{54}^y - r_{54})
\nonumber
\end{eqnarray}

\noindent
\begin{eqnarray}
-\,
\Psi_A(0, \, 0, \, r_{53}^y, \, r_{54}^y),
\label{evolutionEY}
\end{eqnarray}

\noindent
where

\noindent
\begin{eqnarray}
r_{53}^y 
\,=\, 
\frac{\langle p_y \varepsilon  \rangle}{\langle \varepsilon^2 \rangle},
\quad
r_{54}^y 
\,=\, 
-\frac{\langle y \varepsilon  \rangle}{\langle \varepsilon^2 \rangle}.
\label{optSolutionYY}
\end{eqnarray}

\noindent
\begin{eqnarray}
\varepsilon_{\sigma}^2 
\,\leftarrow\, 
\varepsilon_{\sigma}^2
\nonumber
\end{eqnarray}

\noindent
\begin{eqnarray}
+
\Psi_A(r_{51}^{\sigma} - r_{51}, r_{52}^{\sigma} - r_{52}, 
r_{53}^{\sigma} - r_{53}, r_{54}^{\sigma} - r_{54})
\nonumber
\end{eqnarray}

\noindent
\begin{eqnarray}
-
\Psi_A(r_{51}^{\sigma}, r_{52}^{\sigma}, r_{53}^{\sigma}, r_{54}^{\sigma}),
\label{evolutionEZ}
\end{eqnarray}

\noindent
where

\noindent
\begin{eqnarray}
A
\left(
\begin{array}{c}
r_{51}^{\sigma}\\ 
r_{52}^{\sigma}\\ 
r_{53}^{\sigma}\\ 
r_{54}^{\sigma}
\end{array}
\right)
=
\langle \sigma \varepsilon \rangle 
\left(
\begin{array}{c}
\langle x \varepsilon \rangle\\
\langle p_x \varepsilon \rangle\\
\langle y \varepsilon \rangle\\
\langle p_y \varepsilon \rangle
\end{array}
\right)
-
\langle \varepsilon^2 \rangle 
\left(
\begin{array}{c}
\langle x \sigma \rangle\\
\langle p_x \sigma \rangle\\
\langle y \sigma \rangle\\
\langle p_y \sigma \rangle
\end{array}
\right).
\label{optSolutionZ}
\end{eqnarray}

\noindent
One sees from the propagation rules obtained that while the beam tilts 
and the beam energy chirp can influence the evolution of the longitudinal 
projected emittance $\varepsilon_{\sigma}$  through the solution of 
the equation (\ref{optSolutionZ}), 
they do not enter the formulas for 
the evolution of the transverse projected emittances
$\varepsilon_x$ and $\varepsilon_y$ at all.

\subsection{Transformation of 2D transverse projected emittance
and transverse coupling terms}

In the previous subsection the evolution of all three 1D projected 
emittances was expressed using the single quadratic form $\Psi_A$.
Unfortunately, to describe the evolution of the 2D transverse 
projected emittance
another, different from $\Psi_A$, quadratic form is needed.
We denote this form $\Psi_B$ and associated it  
with the positive definite symmetric matrix

\noindent
\begin{eqnarray}
B
=
\left|
\left(
-J_6 \Sigma J_6
\right)\left\{1,2,3,4,5\right\}
\right|
\nonumber
\end{eqnarray}

\noindent
\begin{eqnarray}
\cdot
\big[
\left(
-J_6 \Sigma J_6
\right)\left\{1,2,3,4,5\right\}
\big]^{-1}\left\{1,2,3,4\right\}.
\label{matrixB}
\end{eqnarray}

\noindent
With the help of this new quadratic form the evolution of the 
2D transverse projected emittance can be expressed as follows:

\noindent
\begin{eqnarray}
\varepsilon_t^2 
\leftarrow
\varepsilon_t^2
\nonumber
\end{eqnarray}

\noindent
\begin{eqnarray}
+\,
\Psi_B(r_{51}^x - r_{51}, r_{52}^x - r_{52}, 
r_{53}^y - r_{53}, r_{54}^y - r_{54})
\nonumber
\end{eqnarray}

\noindent
\begin{eqnarray}
-\,
\Psi_B(r_{51}^x, r_{52}^x, r_{53}^y, r_{54}^y),
\label{evolutionET}
\end{eqnarray}

\noindent
where $r_{51}^x$, $r_{52}^x$, $r_{53}^y$, and $r_{54}^y$
are the same as given by the 
formulas (\ref{optSolutionXX}) and (\ref{optSolutionYY}).

Note that though one may accept without additional questions the fact that 
the right hand sides of the formulas (\ref{evolutionEX}) and (\ref{evolutionEY})
are the second order polynomials with respect to the lattice dispersions,
the same property of the right hand side of the formula (\ref{evolutionET}) 
might be somewhat more surprising. For example, let us assume that the beam matrix
is transversely uncoupled at the exit of the dispersive part of the correction system.
Then the right hand side of the formula (\ref{evolutionET}) must coincide with 
the product of the right hand sides of the formulas (\ref{evolutionEX}) 
and (\ref{evolutionEY}) and therefore should contain a polynomial of the fourth order 
with respect to the variables $r_{51}$, $r_{52}$, $r_{53}$, and $r_{54}$.
Because the formula (\ref{evolutionET}) does not provide such a possibility, 
our assumption must be wrong and, during the passage of the dispersive part 
of the correction system, the coupling between transverse degrees of freedom in the
beam matrix must be created. This coupling is described by the following 
propagation rules

\noindent
\begin{eqnarray}
\langle x y  \rangle 
\leftarrow
\langle x y  \rangle 
+
\langle y \varepsilon \rangle
r_{52} 
+
\langle x \varepsilon \rangle
r_{54} 
+
\langle \varepsilon^2 \rangle
r_{52} r_{54},
\label{CoplingXY}
\end{eqnarray}

\noindent
\begin{eqnarray}
\langle x p_y  \rangle 
\leftarrow
\langle x p_y  \rangle 
+
\langle p_y \varepsilon \rangle
r_{52} 
-
\langle x \varepsilon \rangle
r_{53} 
-
\langle \varepsilon^2 \rangle
r_{52} r_{53},
\label{CoplingXPY}
\end{eqnarray}

\noindent
\begin{eqnarray}
\langle y p_x  \rangle 
\leftarrow
\langle y p_x  \rangle 
-
\langle y \varepsilon \rangle
r_{51} 
+
\langle p_x \varepsilon \rangle
r_{54} 
-
\langle \varepsilon^2 \rangle
r_{51} r_{54},
\label{CoplingYPX}
\end{eqnarray}

\noindent
\begin{eqnarray}
\langle p_x p_y  \rangle 
\leftarrow
\langle p_x p_y  \rangle 
-
\langle p_y \varepsilon \rangle
r_{51} 
-
\langle p_x \varepsilon \rangle
r_{53} 
+
\langle \varepsilon^2 \rangle
r_{51} r_{53},
\label{CoplingPXPY}
\end{eqnarray}

\noindent
and, as it can be shown by direct calculations, it really does not allow to 
the terms of the order higher than two with respect to the variables 
$r_{51}$, $r_{52}$, $r_{53}$, and $r_{54}$
to appear in the right hand side of the formula (\ref{evolutionET}).

\subsection{Transformation of transverse to longitudinal coupling terms}

Transformation of the transverse to longitudinal coupling terms
in accordance with the transport rule (\ref{S01})
produces the following changes in the beam dispersions 

\noindent
\begin{eqnarray}
\left(
\begin{array}{c}
\langle x   \varepsilon \rangle\\
\langle p_x \varepsilon \rangle\\
\langle y   \varepsilon \rangle\\
\langle p_y \varepsilon \rangle
\end{array}
\right)
\leftarrow
\left(
\begin{array}{c}
\langle x   \varepsilon \rangle\\
\langle p_x \varepsilon \rangle\\
\langle y   \varepsilon \rangle\\
\langle p_y \varepsilon \rangle
\end{array}
\right)
+
\langle \varepsilon^2 \rangle
J_4
\left(
\begin{array}{c}
r_{51}\\
r_{52}\\
r_{53}\\
r_{54}
\end{array}
\right),
\label{evolutionBD}
\end{eqnarray}

\noindent
and the following changes in the beam tilts

\noindent
\begin{eqnarray}
\left(
\begin{array}{c}
\langle x   \sigma \rangle\\
\langle p_x \sigma \rangle\\
\langle y   \sigma \rangle\\
\langle p_y \sigma \rangle
\end{array}
\right)
\leftarrow
\left(
\begin{array}{c}
\langle x   \sigma \rangle\\
\langle p_x \sigma \rangle\\
\langle y   \sigma \rangle\\
\langle p_y \sigma \rangle
\end{array}
\right)
+
r_{56}
\left(
\begin{array}{c}
\langle x   \varepsilon \rangle\\
\langle p_x \varepsilon \rangle\\
\langle y   \varepsilon \rangle\\
\langle p_y \varepsilon \rangle
\end{array}
\right)
\nonumber
\end{eqnarray}

\noindent
\begin{eqnarray}
+
\left(
\Sigma\left\{1,2,3,4\right\}
+
\lambda\,
J_4
\right)\,
\left(
\begin{array}{c}
r_{51}\\
r_{52}\\
r_{53}\\
r_{54}
\end{array}
\right),
\label{evolutionBT}
\end{eqnarray}

\noindent
where the parameter $\lambda$ is defined by the expression

\noindent
\begin{eqnarray}
\lambda =
\left(
\begin{array}{c}
\langle x   \varepsilon \rangle\\
\langle p_x \varepsilon \rangle\\
\langle y   \varepsilon \rangle\\
\langle p_y \varepsilon \rangle
\end{array}
\right)^{\top}
\left(
\begin{array}{c}
r_{51}\\
r_{52}\\
r_{53}\\
r_{54}
\end{array}
\right)
+
\langle \sigma   \varepsilon \rangle
+
r_{56}
\langle \varepsilon^2 \rangle.
\label{LambdaD}
\end{eqnarray}

\subsection{Transformation of longitudinal moments}

The transformation of the beam energy chirp 
$\langle \sigma \varepsilon  \rangle$
is given by the above introduced parameter $\lambda$

\noindent
\begin{eqnarray}
\langle \sigma \varepsilon  \rangle 
\,\leftarrow\,
\lambda,
\label{EnergyChirp}
\end{eqnarray}

\noindent
and for the description of the change in the rms
bunch length squared $\langle \sigma^2 \rangle$
the new quadratic form is needed again. 
This time it must be quadratic form not in four but
in five variables, because as far as the evolution of 
the projected emittances does not depend from the $r_{56}$ 
matrix coefficient, the evolution of the bunch length 
certainly does. So, let us introduce quadratic form $\Psi_E$ 
associated with the positive definite
matrix 

\noindent
\begin{eqnarray}
E \,=\, \Sigma\left\{1,2,3,4,6\right\}
\label{matrE}
\end{eqnarray}

\noindent
and represent the evolution of $\langle \sigma^2 \rangle$ in the form

\noindent
\begin{eqnarray}
\langle \sigma^2 \rangle 
\,\leftarrow\, 
\langle \sigma^2 \rangle 
+
\Psi_E(r_{51}^{s} - r_{51}, r_{52}^{s} - r_{52}, 
\nonumber
\end{eqnarray}

\noindent
\begin{eqnarray}
r_{53}^{s} - r_{53}, r_{54}^{s} - r_{54}, r_{56}^{s} - r_{56})
\nonumber
\end{eqnarray}

\noindent
\begin{eqnarray}
-
\Psi_E(r_{51}^{s}, r_{52}^{s}, r_{53}^{s}, r_{54}^{s}, r_{56}^{s}),
\label{evolutionBL11}
\end{eqnarray}

\noindent
where

\noindent
\begin{eqnarray}
\Sigma\left\{1,2,3,4,6\right\}
\left(
\begin{array}{c}
r_{51}^{s}\\ 
r_{52}^{s}\\ 
r_{53}^{s}\\ 
r_{54}^{s}\\
r_{56}^{s}
\end{array}
\right)
=
-
\left(
\begin{array}{c}
\langle  x     \sigma \rangle\\
\langle  p_x   \sigma \rangle\\
\langle  y     \sigma \rangle\\
\langle  p_y   \sigma \rangle\\
\langle \sigma \varepsilon \rangle
\end{array}
\right).
\label{optSolutionBL11}
\end{eqnarray}

Note that, if for some reasons the variation of the $r_{56}$ coefficient
is not allowed and it can be treated as a given parameter, then one can 
return to the usage of quadratic form in four variables and
rearrange the formulas (\ref{evolutionBL11}) and (\ref{optSolutionBL11}) 
as follows:

\noindent
\begin{eqnarray}
\langle \sigma^2 \rangle 
\,\leftarrow\, 
\langle \sigma^2 \rangle 
+
2 \,\langle \sigma \varepsilon \rangle\,
r_{56}
+
\langle \varepsilon^2 \rangle\,
r_{56}^2
\nonumber
\end{eqnarray}

\noindent
\begin{eqnarray}
+
\Psi_E(\tilde{r}_{51}^{s} - r_{51}, \tilde{r}_{52}^{s} - r_{52}, 
\tilde{r}_{53}^{s} - r_{53}, \tilde{r}_{54}^{s} - r_{54}, 0)
\nonumber
\end{eqnarray}

\noindent
\begin{eqnarray}
-
\Psi_E(\tilde{r}_{51}^{s}, \tilde{r}_{52}^{s}, 
\tilde{r}_{53}^{s}, \tilde{r}_{54}^{s}, 0),
\label{evolutionBL11003}
\end{eqnarray}

\noindent
where now

\noindent
\begin{eqnarray}
\Sigma\left\{1,2,3,4\right\}
\left(
\begin{array}{c}
\tilde{r}_{51}^{s}\\ 
\tilde{r}_{52}^{s}\\ 
\tilde{r}_{53}^{s}\\ 
\tilde{r}_{54}^{s}
\end{array}
\right)
\nonumber
\end{eqnarray}

\noindent
\begin{eqnarray}
=
-
\left(
\begin{array}{c}
\langle x \sigma \rangle\\
\langle p_x \sigma \rangle\\
\langle y \sigma \rangle\\
\langle p_y \sigma \rangle
\end{array}
\right)
-
r_{56}
\left(
\begin{array}{c}
\langle x \varepsilon \rangle\\
\langle p_x \varepsilon \rangle\\
\langle y \varepsilon \rangle\\
\langle p_y \varepsilon \rangle
\end{array}
\right).
\label{optSolutionBL967}
\end{eqnarray}

\section{Optimal solution for minimization of transverse projected 
emittances and its properties}

With the formulas developed in the previous section for the emittance transport 
the problem of optimization of transverse projected emittances by an appropriate
choice of the lattice dispersions becomes (at least from the theoretical
point of view) fairly simple and straightforward.
For example, the formula (\ref{evolutionEX})
tell us that the change in the horizontal projected emittance $\varepsilon_x$
after the system passage is the same for 
all lattice dispersions $r_{51}$ and $r_{52}$ belonging to 
the same level set 

\noindent
\begin{eqnarray}
\Psi_A(r_{51}^x - r_{51}, r_{52}^x - r_{52}, 0, 0) = const \geq 0.
\label{ellipseEX01}
\end{eqnarray}

\noindent
Because the function $\Psi_A$ is a positive definite quadratic form
its level sets for $const > 0$ are ellipses all centered at the same point

\noindent
\begin{eqnarray}
r_{51} = r_{51}^x, 
\quad
r_{52} = r_{52}^x
\label{ellipseEXC}
\end{eqnarray}

\noindent
and contracting to this point as $const \rightarrow 0$. The level set

\noindent
\begin{eqnarray}
\Psi_A(r_{51}^x - r_{51}, r_{52}^x - r_{52}, 0, 0)
=
\Psi_A(r_{51}^x, r_{52}^x, 0, 0)
\label{ellipseEX}
\end{eqnarray}

\noindent
plays a special role. It separates the lattice dispersions which lead 
to the emittance increase from the lattice dispersions which provide emittance 
reduction or preservation. The level surface (\ref{ellipseEX}) is an ellipse 
if at least one horizontal beam dispersion is nonzero at the correction system 
entrance and it is a point coinciding with the common center (\ref{ellipseEXC})
of all ellipses (\ref{ellipseEX01}) otherwise. In any case there exists unique 
optimal choice (optimal solution) for the horizontal lattice dispersions which
is given by the equation (\ref{ellipseEXC}) and which provides the largest possible 
reduction of the horizontal projected emittance $\varepsilon_x$ (the largest 
possible reduction is zero if both horizontal beam dispersions are equal to zero).

By analogy, the optimal solution for the transport of the vertical 
projected emittance $\varepsilon_y$ is reached in the point

\noindent
\begin{eqnarray}
r_{53} = r_{53}^y, 
\quad
r_{54} = r_{54}^y, 
\label{ellipseEYC}
\end{eqnarray}

\noindent
and the optimal solution for the transport of the complete 2D transverse 
projected emittance $\varepsilon_t$ is given by the union
of the solution for the horizontal motion (\ref{ellipseEXC}) and 
the solution for the vertical motion (\ref{ellipseEYC}),
which is a very pleasant fact (in general, if the chosen lattice dispersions
are different from the optimal solution, then the reduction of 
both $\varepsilon_x$ and $\varepsilon_y$ does not guarantee the reduction 
of $\varepsilon_t$, and vice versa).

One sees that the values of the lattice dispersions required for the simultaneous 
minimization of all transverse projected emittances are completely determined 
by the values of the beam dispersions and the beam rms energy spread, but, even
if these quantities are unknown and there is no appropriate diagnostics
to measure them, the projected emittances still can be optimized if there is
a possibility to measure the horizontal and vertical projected emittances
downstream of the correction system. In this situation, minimization of
emittances can be done iteratively (and independently for horizontal and
vertical degrees of freedom) employing one of the many available
effective algorithms for minimizing a convex quadratic objective function
of two variables.

\subsection{Effect of the optimal solution on the beam transport}

The optimal solution for all four lattice dispersions
$r_{51}$, $r_{52}$, $r_{53}$, and $r_{54}$ can be written in the form

\noindent
\begin{eqnarray}
\left(
\begin{array}{c}
r_{51}\\
r_{52}\\
r_{53}\\
r_{54}
\end{array}
\right)
\,=\,
\frac{1}{\langle \varepsilon^2 \rangle}
J_4
\left(
\begin{array}{c}
\langle x   \varepsilon \rangle\\
\langle p_x \varepsilon \rangle\\
\langle y   \varepsilon \rangle\\
\langle p_y \varepsilon \rangle
\end{array}
\right)
\label{OptSol01}
\end{eqnarray}

\noindent
and therefore satisfies the orthogonality condition

\noindent
\begin{eqnarray}
\left(
\begin{array}{c}
\langle x   \varepsilon \rangle\\
\langle p_x \varepsilon \rangle\\
\langle y   \varepsilon \rangle\\
\langle p_y \varepsilon \rangle
\end{array}
\right)^{\top}
\left(
\begin{array}{c}
r_{51}\\
r_{52}\\
r_{53}\\
r_{54}
\end{array}
\right)
= 0.
\label{OptSol02}
\end{eqnarray}

With the optimal choice of the lattice dispersions (\ref{OptSol01}) 
the beam dispersions are zeroed at the correction system exit and 
the tilts are transformed according to the rule

\noindent
\begin{eqnarray}
\left(
\begin{array}{c}
\langle x \sigma \rangle\\
\langle p_x \sigma \rangle\\
\langle y \sigma \rangle\\
\langle p_y \sigma \rangle
\end{array}
\right)
\leftarrow
\left(
\begin{array}{c}
\langle x \sigma \rangle\\
\langle p_x \sigma \rangle\\
\langle y \sigma \rangle\\
\langle p_y \sigma \rangle
\end{array}
\right)
\nonumber
\end{eqnarray}

\noindent
\begin{eqnarray}
+
\frac{
\Sigma\left\{1,2,3,4\right\}
J_4
-
\langle \sigma \varepsilon \rangle
I_4
}
{\langle \varepsilon^2 \rangle}
\left(
\begin{array}{c}
\langle x \varepsilon  \rangle\\
\langle p_x \varepsilon  \rangle\\
\langle y \varepsilon  \rangle\\
\langle p_y \varepsilon  \rangle
\end{array}
\right),
\label{bt0003}
\end{eqnarray}

\noindent
where, as one sees, the dependence from the $r_{56}$ matrix coefficient 
presented in the formula (\ref{evolutionBT}) disappeared, though 
no assumptions were made about this coefficient and the optimal lattice 
dispersions (\ref{OptSol01}) also do not depend on it.

The transport of the transverse coupling terms is given now by the formulas

\noindent
\begin{eqnarray}
\langle x y  \rangle 
\leftarrow
\langle x y  \rangle 
-
\frac{\langle x \varepsilon \rangle
\langle y \varepsilon \rangle}{\langle \varepsilon^2 \rangle},
\label{CoplingXYOpt}
\end{eqnarray}

\noindent
\begin{eqnarray}
\langle x p_y  \rangle 
\leftarrow
\langle x p_y  \rangle 
-
\frac{\langle x \varepsilon \rangle
\langle p_y \varepsilon \rangle}{\langle \varepsilon^2 \rangle},
\label{CoplingXPYOpt}
\end{eqnarray}

\noindent
\begin{eqnarray}
\langle y p_x  \rangle 
\leftarrow
\langle y p_x \rangle 
-
\frac{\langle y \varepsilon \rangle
\langle p_x \varepsilon \rangle}{\langle \varepsilon^2 \rangle},
\label{CoplingYPXOpt}
\end{eqnarray}

\noindent
\begin{eqnarray}
\langle p_x p_y  \rangle 
\leftarrow
\langle p_x p_y  \rangle 
-
\frac{\langle p_x \varepsilon \rangle
\langle p_y \varepsilon \rangle}{\langle \varepsilon^2 \rangle},
\label{CoplingPXPYOpt}
\end{eqnarray}

\noindent
and one sees that if both, horizontal and vertical, beam dispersion vectors
are nonzero at the entrance, then the interplay between them during the passage 
of the dispersive part of the correction system becomes a source of the 
transverse coupling. This coupling could be removed by adding to the correction
system an appropriate set of the skew quadrupoles and, therefore, the 1D transverse
projected emittances can be reduced even further. 

In order to find better expressions for the reduction of the transverse 
projected emittances than the expressions which one can obtain by the direct 
substitution of the optimal solution (\ref{OptSol01}) into the emittance 
propagation formulas, let us introduce positive definite quadratic forms 
$\Psi_C$ and $\Psi_D$ associated with the positive definite matrices

\noindent
\begin{eqnarray}
C
=
\left|
\Sigma\left\{1,2,3,4\right\}
\right|
\left(
\Sigma\left\{1,2,3,4\right\}
\right)^{-1}
\label{matrixC}
\end{eqnarray}

\noindent
and

\noindent
\begin{eqnarray}
D
=
-J_4 \left(\Sigma\left\{1,2,3,4\right\}\right) J_4,
\label{matrixD}
\end{eqnarray}

\noindent
respectively. The advantage of these quadratic forms over the quadratic 
forms $\Psi_A$ and $\Psi_B$ is that the elements of their matrices $C$ 
and $D$ are functions of the transverse beam moments only and do not depend 
on the beam moments involving the longitudinal variable $\varepsilon$ as do 
the elements of the matrices $A$ and $B$. Besides that 

\noindent
\begin{eqnarray}
\Psi_D(u_1, \,u_2,\, 0,\, 0) 
\,=\, I_{cs}^x(u_1, \, u_2),
\label{tmpCS01}
\end{eqnarray}

\noindent
\begin{eqnarray}
\Psi_D(0, \, 0, \, u_3, \, u_4) 
\,=\, I_{cs}^y(u_3, \, u_4),
\label{tmpCS02}
\end{eqnarray}

\noindent
where

\noindent
\begin{eqnarray}
I_{cs}^x(u_1, \, u_2)
=
\langle p_x^2  \rangle \, u_1^2
- 
2 \, \langle x p_x  \rangle \, u_1 u_2
+ 
\langle x^2  \rangle \, u_2^2,
\label{noCouplXY3}
\end{eqnarray}

\noindent
\begin{eqnarray}
I_{cs}^y(u_3, u_4)
=
\langle p_y^2  \rangle \, u_3^2
- 
2 \, \langle y p_y  \rangle \, u_3 u_4
+ 
\langle y^2  \rangle \, u_4^2
\label{noCouplXY4}
\end{eqnarray}

\noindent
are the familiar (but nonnormalized) horizontal and vertical 
Courant-Snyder quadratic forms.

With the help of the quadratic forms $\Psi_C$ and $\Psi_D$ the evolution
of the transverse projected emittances for the optimal choice of the
lattice dispersions can be expressed as follows:

\noindent
\begin{eqnarray}
\varepsilon_x^2 
\,\leftarrow\,
\varepsilon_x^2
\,-\, 
\frac{1}{\langle \varepsilon^2 \rangle}
\cdot
\Psi_D(
\langle x \varepsilon \rangle, \,
\langle p_x \varepsilon \rangle, \,
0, \,0)
\nonumber
\end{eqnarray}

\noindent
\begin{eqnarray}
=
\varepsilon_x^2
\,-\, 
\frac{1}{\langle \varepsilon^2 \rangle}
\cdot
I_{cs}^x(
\langle x \varepsilon \rangle, \,
\langle p_x \varepsilon \rangle),
\label{evolutionEXOPT}
\end{eqnarray}

\noindent
\begin{eqnarray}
\varepsilon_y^2 
\,\leftarrow\,
\varepsilon_y^2
\,-\, 
\frac{1}{\langle \varepsilon^2 \rangle}
\cdot
\Psi_D(
0, \,
0, \,
\langle y \varepsilon \rangle, \,
\langle p_y \varepsilon \rangle)
\nonumber
\end{eqnarray}

\noindent
\begin{eqnarray}
=
\varepsilon_y^2
\,-\, 
\frac{1}{\langle \varepsilon^2 \rangle}
\cdot
I_{cs}^y(
\langle y \varepsilon \rangle, \,
\langle p_y \varepsilon \rangle),
\label{evolutionEYOPT}
\end{eqnarray}

\noindent
\begin{eqnarray}
\varepsilon_t^2 
\,\leftarrow\,
\varepsilon_t^2
\,-\, 
\frac{1}{\langle \varepsilon^2 \rangle}
\cdot
\Psi_C(
\langle x \varepsilon \rangle, \,
\langle p_x \varepsilon \rangle, \,
\langle y \varepsilon \rangle, \,
\langle p_y \varepsilon \rangle),
\label{evolutionETOPT}
\end{eqnarray}

\noindent
and for the longitudinal projected emittance one obtains:

\noindent
\begin{eqnarray}
\varepsilon_{\sigma}^2 
\leftarrow 
\varepsilon_{\sigma}^2
+
\frac{1}{\langle \varepsilon^2 \rangle}
\cdot
\left[
\Psi_D(
d_{x}^{\sigma}   - \langle x   \varepsilon \rangle, 
d_{p_x}^{\sigma} - \langle p_x \varepsilon \rangle, 
\right.
\nonumber
\end{eqnarray}

\noindent
\begin{eqnarray}
\left.
d_{y}^{\sigma}   - \langle y   \varepsilon \rangle, 
d_{p_y}^{\sigma} - \langle p_y \varepsilon \rangle
)
-
\Psi_D(
d_{x}^{\sigma},   
d_{p_x}^{\sigma}, 
d_{y}^{\sigma},  
d_{p_y}^{\sigma})
\right],
\label{evolutionEZOPT}
\end{eqnarray}

\noindent
where

\noindent
\begin{eqnarray}
\left(
\begin{array}{c}
d_{x}^{\sigma}\\ 
d_{p_x}^{\sigma}\\ 
d_{y}^{\sigma}\\ 
d_{p_y}^{\sigma}
\end{array}
\right)
=
\langle \varepsilon^2 \rangle
J_4 
\left(
\Sigma\left\{1,2,3,4\right\}
\right)^{-1}
\left(
\begin{array}{c}
\langle x \sigma \rangle\\
\langle p_x \sigma \rangle\\
\langle y \sigma \rangle\\
\langle p_y \sigma \rangle
\end{array}
\right).
\label{optSolutionZOPT2}
\end{eqnarray}

\noindent
One sees that the influence of the energy chirp $\langle \sigma \varepsilon \rangle$ 
on the propagation  of the longitudinal projected emittance, which was presented in 
the formula (\ref{evolutionEZ}) through the solution of the equation (\ref{optSolutionZ}), 
is now canceled. As concerning the transport of the energy chirp itself, it is
simplified owing to the orthogonality condition (\ref{OptSol02}) to the form

\noindent
\begin{eqnarray}
\langle \sigma \varepsilon  \rangle 
\,\leftarrow\,
\langle \sigma \varepsilon  \rangle 
+
r_{56}
\langle \varepsilon^2 \rangle,
\label{EnergyChirpOPT}
\end{eqnarray}

\noindent
and the propagation formula for the rms bunch length squared $\langle \sigma^2 \rangle$, 
if needed, can be obtained using the equations (\ref{evolutionEZOPT}) 
and (\ref{EnergyChirpOPT}), and the relation

\noindent
\begin{eqnarray}
\langle \sigma^2 \rangle
\,=\,
\frac{\varepsilon_{\sigma}^2 \,+\, \langle \sigma \varepsilon \rangle^2}
{\langle \varepsilon^2 \rangle}. 
\label{BLS2}
\end{eqnarray}

\subsection{Transversely uncoupled beam at the correction
system entrance}

The formulas (\ref{evolutionETOPT}) and (\ref{evolutionEZOPT}) 
for the transport of the 2D transverse projected emittance $\varepsilon_t$ 
and the longitudinal projected emittance $\varepsilon_{\sigma}$ can be
further simplified if one assumes that the conditions (\ref{noCouplXY678}) 
hold and the transverse degrees of freedom in the beam matrix $\Sigma$ 
are decoupled from each other at the correction system entrance. 
With this assumption matrices $C$ and $D$ become block diagonal,
quadratic forms $\Psi_C$ and $\Psi_D$ get representations

\noindent
\begin{eqnarray}
\Psi_D(u_1, u_2, u_3, u_4)
=
I_{cs}^x(u_1, u_2)
+
I_{cs}^y(u_3, u_4),
\label{noCouplXY21}
\end{eqnarray}

\noindent
\begin{eqnarray}
\Psi_C(u_1, u_2, u_3, u_4)
=
\varepsilon_y^2
I_{cs}^x(u_1, u_2)
+
\varepsilon_x^2
I_{cs}^y(u_3, u_4),
\label{noCouplXY22}
\end{eqnarray}

\noindent
and, as the result, one obtains

\noindent
\begin{eqnarray}
\varepsilon_t^2 
\,\leftarrow\,
\varepsilon_t^2
\nonumber
\end{eqnarray}

\noindent
\begin{eqnarray}
-\, 
\frac{1}{\langle \varepsilon^2 \rangle}
\cdot
\left[
\varepsilon_y^2 
I_{cs}^x\left(
\langle x \varepsilon \rangle, 
\langle p_x \varepsilon \rangle\right)
+
\varepsilon_x^2 
I_{cs}^y\left(
\langle y \varepsilon \rangle, 
\langle p_y \varepsilon \rangle\right)
\right],
\label{evolutionETOPT2}
\end{eqnarray}

\noindent
\begin{eqnarray}
\varepsilon_{\sigma}^2 
\,\leftarrow\,
\varepsilon_{\sigma}^2
\nonumber
\end{eqnarray}

\noindent
\begin{eqnarray}
+
\frac{1}{\langle \varepsilon^2 \rangle}
I_{cs}^x(
\langle x \varepsilon \rangle, 
\langle p_x \varepsilon \rangle)
+
\frac{1}{\langle \varepsilon^2 \rangle}
I_{cs}^y(
\langle y \varepsilon \rangle, 
\langle p_y \varepsilon \rangle)
\nonumber
\end{eqnarray}

\noindent
\begin{eqnarray}
+
2
\left|
\begin{array}{cc}
\langle x   \sigma \rangle & \langle x   \varepsilon \rangle\\
\langle p_x \sigma \rangle & \langle p_x \varepsilon \rangle 
\end{array}
\right|
+
2
\left|
\begin{array}{cc}
\langle y   \sigma \rangle & \langle y   \varepsilon \rangle\\
\langle p_y \sigma \rangle & \langle p_y \varepsilon \rangle 
\end{array}
\right|.
\label{optLongEmmNoCoupl}
\end{eqnarray}

\noindent
Note that the formula (\ref{optLongEmmNoCoupl}) also can be obtained from equations 
(\ref{CoplingXYOpt})-(\ref{CoplingPXPYOpt}), (\ref{evolutionEXOPT})-(\ref{evolutionEYOPT})
and conditions (\ref{noCouplXY678}) using conservation 
of the Lysenko invariant \cite{Lysenko1,Lysenko2} 

\noindent
\begin{eqnarray}
I_{ls} 
= \varepsilon_x^2
+ \varepsilon_y^2
+ \varepsilon_{\sigma}^2
+
2
\left|
\begin{array}{cc}
\langle x y   \rangle & \langle x p_y   \rangle\\
\langle y p_x \rangle & \langle p_x p_y \rangle 
\end{array}
\right|
\nonumber
\end{eqnarray}

\noindent
\begin{eqnarray}
+
2
\left|
\begin{array}{cc}
\langle x   \sigma \rangle & \langle x   \varepsilon \rangle\\
\langle p_x \sigma \rangle & \langle p_x \varepsilon \rangle 
\end{array}
\right|
+
2
\left|
\begin{array}{cc}
\langle y   \sigma \rangle & \langle y   \varepsilon \rangle\\
\langle p_y \sigma \rangle & \langle p_y \varepsilon \rangle 
\end{array}
\right|
\label{LysenkoInvariant}
\end{eqnarray}

\noindent
during symplectic transport of the beam matrix.

\section{Miscellaneous}

\subsection{Optimization of longitudinal projected emittance}

As the formulas (\ref{evolutionEZ}) and (\ref{optSolutionZ}) show, the problem 
of optimization of the longitudinal projected emittance by the proper choice 
of the lattice dispersions (when considered alone) has the same geometry as 
the corresponding problems for the transverse projected emittances. 
The resulting $\varepsilon_{\sigma}$ increase or reduction depends 
on the positioning of the chosen lattice dispersions $r_{51}$, $r_{52}$, $r_{53}$, 
and $r_{54}$ with respect to the four dimensional ellipsoid 

\noindent
\begin{eqnarray}
\Psi_A(r_{51}^{\sigma} - r_{51}, r_{52}^{\sigma} - r_{52}, 
r_{53}^{\sigma} - r_{53}, r_{54}^{\sigma} - r_{54})
\nonumber
\end{eqnarray}

\noindent
\begin{eqnarray}
=
\Psi_A(r_{51}^{\sigma}, r_{52}^{\sigma}, r_{53}^{\sigma}, r_{54}^{\sigma}),
\label{evolutionEZLS}
\end{eqnarray}

\noindent
and the optimal solution is obviously reached in the point

\noindent
\begin{eqnarray}
r_{51} = r_{51}^{\sigma}, \quad
r_{52} = r_{52}^{\sigma}, \quad
r_{53} = r_{53}^{\sigma}, \quad
r_{54} = r_{54}^{\sigma}.
\label{evolutionEZLS2}
\end{eqnarray}

\noindent
But, in contrast to the transport of the transverse projected emittances,
the longitudinal projected emittance can be reduced even when
all beam dispersions are equal to zero at the correction system entrance, 
because (according to the equation (\ref{optSolutionZ})) the optimal 
solution (\ref{evolutionEZLS2}) depends also on the values of the beam tilts.

It is outside of the purpose of this paper to make a detailed study of the influence 
of the solution (\ref{evolutionEZLS2}) on the propagation of the other beam parameters 
and let us only note that it makes the vector of the beam dispersions and the vector 
of the beam tilts linearly dependent (parallel) at the exit of the dispersive part of 
the correction system. It comes from the fact that the choice of the lattice dispersions 
according to the equations (\ref{evolutionEZLS2}) gives us the following transport rule
for the beam dispersions

\noindent
\begin{eqnarray}
\left(
\begin{array}{c}
\langle x   \varepsilon \rangle\\
\langle p_x \varepsilon \rangle\\
\langle y   \varepsilon \rangle\\
\langle p_y \varepsilon \rangle
\end{array}
\right)
\leftarrow
\left(
\begin{array}{c}
\langle x   \varepsilon \rangle\\
\langle p_x \varepsilon \rangle\\
\langle y   \varepsilon \rangle\\
\langle p_y \varepsilon \rangle
\end{array}
\right)
+
\langle \varepsilon^2 \rangle
J_4
\left(
\begin{array}{c}
r_{51}^{\sigma}\\
r_{52}^{\sigma}\\
r_{53}^{\sigma}\\
r_{54}^{\sigma}
\end{array}
\right),
\label{evolutionBDoptL}
\end{eqnarray}

\noindent
and the following transport rule for the beam tilts

\noindent
\begin{eqnarray}
\left(
\begin{array}{c}
\langle x   \sigma \rangle\\
\langle p_x \sigma \rangle\\
\langle y   \sigma \rangle\\
\langle p_y \sigma \rangle
\end{array}
\right)
\leftarrow
\frac{\lambda_{\sigma}}{\langle \varepsilon^2 \rangle}
\left[
\left(
\begin{array}{c}
\langle x   \varepsilon \rangle\\
\langle p_x \varepsilon \rangle\\
\langle y   \varepsilon \rangle\\
\langle p_y \varepsilon \rangle
\end{array}
\right)
+
\langle \varepsilon^2 \rangle
J_4
\left(
\begin{array}{c}
r_{51}^{\sigma}\\
r_{52}^{\sigma}\\
r_{53}^{\sigma}\\
r_{54}^{\sigma}
\end{array}
\right)
\right],
\label{evolutionBToptL}
\end{eqnarray}

\noindent
where the parameter $\lambda_{\sigma}$ is defined by the expression

\noindent
\begin{eqnarray}
\lambda_{\sigma} =
\left(
\begin{array}{c}
\langle x   \varepsilon \rangle\\
\langle p_x \varepsilon \rangle\\
\langle y   \varepsilon \rangle\\
\langle p_y \varepsilon \rangle
\end{array}
\right)^{\top}
\left(
\begin{array}{c}
r_{51}^{\sigma}\\
r_{52}^{\sigma}\\
r_{53}^{\sigma}\\
r_{54}^{\sigma}
\end{array}
\right)
+
\langle \sigma   \varepsilon \rangle
+
r_{56}
\langle \varepsilon^2 \rangle.
\label{LambdaDoptL}
\end{eqnarray}

\noindent
Thus, what is important for our further considerations, both solutions 
(\ref{OptSol01}) and (\ref{evolutionEZLS2}) zero the last two terms
in the Lysenko invariant (\ref{LysenkoInvariant}) at the exit of the 
correction system.

\subsection{Conditions for simultaneous optimization 
of transverse and longitudinal projected emittances}

With our approach, the problem of the simultaneous optimization of the
selected projected emittances by a proper choice of the lattice dispersions
becomes a geometrical problem. For example, emittances $\varepsilon_x$ 
and $\varepsilon_{\sigma}$ can be decreased simultaneously if and only if 
the surfaces (\ref{evolutionEZLS}) and (\ref{ellipseEX}) are non-degenerate
(i.e. they are not points but real ellipsoids)
and the projection of the inner points of the ellipsoid (\ref{evolutionEZLS}) 
onto the plane $r_{53} = r_{54} = 0$ has nonempty 
intersection with the set of the inner points of the ellipse (\ref{ellipseEX}).

The optimal solutions (\ref{OptSol01}) and (\ref{evolutionEZLS2}) 
will be equal to each other and therefore the maximal possible reductions
will be achieved for all, horizontal and vertical, projected emittances
simultaneously, if and only if the following relations between
the elements of the beam matrix $\Sigma$ hold

\noindent
\begin{eqnarray}
\left(
\begin{array}{c}
\langle x \sigma \rangle\\
\langle p_x \sigma \rangle\\
\langle y \sigma \rangle\\
\langle p_y \sigma \rangle
\end{array}
\right)
=
\frac{
\langle \sigma \varepsilon \rangle
I_4
-
\Sigma\left\{1,2,3,4\right\}
J_4
}
{\langle \varepsilon^2 \rangle}
\left(
\begin{array}{c}
\langle x \varepsilon  \rangle\\
\langle p_x \varepsilon  \rangle\\
\langle y \varepsilon  \rangle\\
\langle p_y \varepsilon  \rangle
\end{array}
\right).
\label{decouplCond1}
\end{eqnarray}

\noindent
We will discuss these relations in more detail in the following subsections and 
now let us only point out that the requirement for solution (\ref{OptSol01}) 
to coincide with the first four components of the vector 

\noindent
\begin{eqnarray}
(r_{51}^s, r_{52}^s, r_{53}^s, r_{54}^s, r_{56}^s)^{\top},
\label{vecS2OPT}
\end{eqnarray}

\noindent
which is defined in the equation (\ref{optSolutionBL11}),
gives us the same relation (\ref{decouplCond1}) and also fixes
the choice for the $r_{56}$ coefficient to the value

\noindent
\begin{eqnarray}
r_{56} \,=\, -\frac{\langle \sigma \varepsilon \rangle}{\langle \varepsilon^2 \rangle},
\label{decouplCond3}
\end{eqnarray}

\noindent
which corresponds to the complete chirp removal at the correction system exit.
Note that the choice of the lattice dispersions (including setting of the $r_{56}$ 
coefficient) to be equal to the values (\ref{vecS2OPT}) minimizes the rms bunch 
length squared $\langle \sigma^2 \rangle$ after the correction system passage.

\subsection{Possibilities for zeroing beam tilts}

According to the relations (\ref{evolutionBT}) and (\ref{LambdaD}) the beam 
tilts can be zeroed at the correction system exit by an appropriate choice 
of the correction lattice dispersions if and only if the system of equations

\noindent
\begin{eqnarray}
\left(
\Sigma\left\{1,2,3,4\right\}
+
\lambda\,
J_4
\right)\,
\left(
\begin{array}{c}
r_{51}\\
r_{52}\\
r_{53}\\
r_{54}
\end{array}
\right)
\nonumber
\end{eqnarray}

\noindent
\begin{eqnarray}
=
-
\left(
\begin{array}{c}
\langle x   \sigma \rangle\\
\langle p_x \sigma \rangle\\
\langle y   \sigma \rangle\\
\langle p_y \sigma \rangle
\end{array}
\right)
-
r_{56}
\left(
\begin{array}{c}
\langle x   \varepsilon \rangle\\
\langle p_x \varepsilon \rangle\\
\langle y   \varepsilon \rangle\\
\langle p_y \varepsilon \rangle
\end{array}
\right),
\label{zbt01}
\end{eqnarray}

\noindent
where

\noindent
\begin{eqnarray}
\lambda =
\left(
\begin{array}{c}
\langle x   \varepsilon \rangle\\
\langle p_x \varepsilon \rangle\\
\langle y   \varepsilon \rangle\\
\langle p_y \varepsilon \rangle
\end{array}
\right)^{\top}
\left(
\begin{array}{c}
r_{51}\\
r_{52}\\
r_{53}\\
r_{54}
\end{array}
\right)
+
\langle \sigma   \varepsilon \rangle
+
r_{56}
\langle \varepsilon^2 \rangle
\label{zbt02}
\end{eqnarray}

\noindent
has at least one real solution with respect to the variables
$r_{51}$, $r_{52}$, $r_{53}$, $r_{54}$, and $r_{56}$.
In general, it is a nonlinear system. Nevertheless, as we prove below, 
it always has at least one real solution for every fixed real value of 
the $r_{56}$ coefficient, which therefore can be treated as a parameter.
To show this, let us assume first that $\lambda$ in the system (\ref{zbt01}) 
is not simply a notation introduced for brevity, but is an additional real-valued 
variable, and let us consider an extended system consisting of equations
(\ref{zbt01}) and (\ref{zbt02}). Now we want to apply the method of successive
elimination of variables to the system obtained and with this purpose in mind
let us observe that

\noindent
\begin{eqnarray}
\left|
\Sigma\left\{1,2,3,4\right\} + \lambda J_4
\right|
=
\left|
\Sigma\left\{1,2,3,4\right\} J_4 - \lambda I_4
\right|.
\label{ZBT01}
\end{eqnarray}

\noindent
Because the matrix $\Sigma\left\{1,2,3,4\right\} J_4$
is similar to the non-degenerated skew symmetric matrix

\noindent
\begin{eqnarray}
(\Sigma\left\{1,2,3,4\right\})^{1/2} J_4 
(\Sigma\left\{1,2,3,4\right\})^{1/2}
\label{ZBT11}
\end{eqnarray}

\noindent
which has only pure imaginary nonzero eigenvalues, the right hand side 
of the equality (\ref{ZBT01}) is nonzero for all real values of $\lambda$ 
and therefore the matrix

\noindent
\begin{eqnarray}
\Sigma\left\{1,2,3,4\right\} + \lambda J_4
\label{ZBT12}
\end{eqnarray}

\noindent
is invertible. It means that for every real value of $\lambda$
equations (\ref{zbt01}) can be solved with respect to the 
variables $r_{51}$, $r_{52}$, $r_{53}$, $r_{54}$ and the solution is unique.
Substituting this solution into equation (\ref{zbt02}) and multiplying 
both sides of the result by

\noindent
\begin{eqnarray}
\left|
\Sigma\left\{1,2,3,4\right\} + \lambda J_4
\right|
=
\lambda^4 
\nonumber
\end{eqnarray}

\noindent
\begin{eqnarray}
- \tfrac{1}{2}
\mbox{tr}\left[
\left(
\Sigma\left\{1,2,3,4\right\} J_4
\right)^2
\right]
\lambda^2
+
\left|
\Sigma\left\{1,2,3,4\right\}\right|,
\label{ZBT02}
\end{eqnarray}

\noindent
we obtain the polynomial equation of the fifth degree with respect to the single 
variable $\lambda$ (consistency equation) and, because the order of this equation 
is odd, it always must have at least one real root.

So we proved that the zeroing of the beam tilts by an appropriate choice of the 
correction lattice dispersions is always possible. At least one solution can be found 
for all real values of the $r_{56}$ matrix coefficient and, for the fixed $r_{56}$ 
value, the number of solutions can vary from one to five.

To be more specific, let us consider a numerical example and take 
as a beam matrix the positive definite matrix

\noindent
\begin{eqnarray}
\Sigma =
\left(
\begin{array}{cccccc}
1 & 0 & 0 & 0 & 0 & 2\\
0 & 1 & 0 & 0 & 2 & 0\\
0 & 0 & 1 & 0 & 0 & 0\\
0 & 0 & 0 & 1 & 0 & 0\\
0 & 2 & 0 & 0 & 6 & 0\\
2 & 0 & 0 & 0 & 0 & 6
\end{array}
\right)
\label{numm01}
\end{eqnarray}

\noindent
in which the vertical degree of freedom is decoupled from the two others.
For this matrix $\Sigma$ the solution of the equation (\ref{zbt01})
with $\lambda$ and $r_{56}$ taken as parameters gives

\noindent
\begin{eqnarray}
r_{51} = -\frac{2 (r_{56} - \lambda)}{1 + \lambda^2},
\quad
r_{52} = -\frac{2 (1 + r_{56} \lambda)}{1 + \lambda^2},
\label{numm02}
\end{eqnarray}

\noindent
\begin{eqnarray}
r_{53} \, = \, r_{54} \, = \, 0,
\label{numm022}
\end{eqnarray}

\noindent
and the fifth order consistency equation for the determination of 
the real values of $\lambda$ 

\noindent
\begin{eqnarray}
\left(\lambda^2 + 1 \right)
\left(\lambda^3 - 6 \,r_{56} \,\lambda^2 - 3 \,\lambda - 2\, r_{56}\right) \,=\, 0
\label{numm03678}
\end{eqnarray}

\noindent
reduces to the cubic equation

\noindent
\begin{eqnarray}
\lambda^3 - 6 \,r_{56} \,\lambda^2 - 3 \,\lambda - 2\, r_{56} \,=\, 0.
\label{numm03}
\end{eqnarray}

\noindent
The discriminant of this cubic equation

\noindent
\begin{eqnarray}
\Delta  = 108 \, \left[ 1 - (2\,r_{56})^2 - (2\,r_{56})^4\right]
\label{numm04}
\end{eqnarray}

\noindent
is positive inside the interval

\noindent
\begin{eqnarray}
\left|r_{56}\right| \,<\,
\frac{1}{2}\sqrt{\frac{\sqrt{5}-1}{2}} \, \approx \, 0.393,
\label{numm05}
\end{eqnarray}

\noindent
is equal to zero at the interval endpoints and is negative outside, which corresponds 
to the existence of three, two and one distinct real roots, respectively.

\begin{figure}[!t]
    \centering
    \includegraphics*[width=70mm]{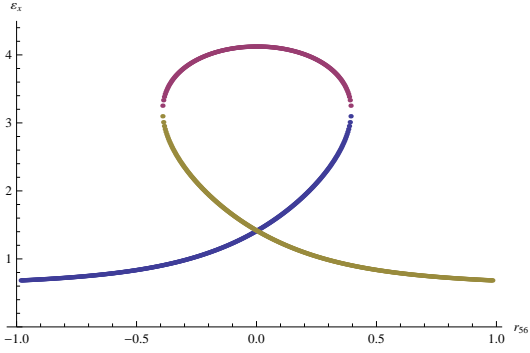}
    \caption{Effect of the beam tilts removal on the horizontal projected 
    emittance $\varepsilon_x$ as a functions of the $r_{56}$ coefficient.}
    \label{fig1}
\end{figure}

The effect of the zeroing of the beam tilts in the matrix (\ref{numm01}) on projected 
emittances is presented at figures 1 and 2, where the resulting emittances are shown
for all possible real solutions of the equation (\ref{numm03}). One has to compare these 
emittances with the emittances $\varepsilon_x = 1$, $\varepsilon_{\sigma} = 6$ of the 
original beam matrix (\ref{numm01}) and with the emittances $\varepsilon_x \approx 0.577$,
$\varepsilon_{\sigma} \approx 5.354$ which can be obtained after removal of the beam dispersions.

\subsection{Conditions for complete transverse to longitudinal decoupling}

The example considered in the previous subsection tells us that zeroing of the beam tilts does 
not necessarily implies reduction of the transverse projected emittances. The situation, 
of course, will be different if zeroing of the beam tilts will simultaneously remove the 
beam dispersions, i.e. if the longitudinal and transverse degrees of freedom in the beam 
matrix $\Sigma$ will be decoupled from each other at the correction system exit. The necessary 
and sufficient conditions for the complete transverse to longitudinal decoupling can be 
obtained by the requirement that the solution for the lattice dispersions (\ref{OptSol01}) 
which removes the beam dispersions also zeros the beam tilts, and substituting (\ref{OptSol01}) 
into the equations (\ref{zbt01}) and (\ref{zbt02}) we obtain (without big surprise) again 
the equations (\ref{decouplCond1}), which were derived as conditions for the simultaneous 
minimization of all projected emittances.

\begin{figure}[!t]
    \centering
    \includegraphics*[width=70mm]{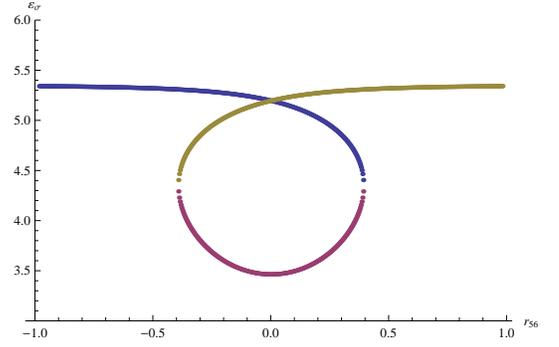}
    \caption{Effect of the beam tilts removal on the longitudinal projected 
    emittance $\varepsilon_{\sigma}$ as a functions of the $r_{56}$ coefficient.}
    \label{fig2}
\end{figure}

The most difficult question, for which we do not have any good answers yet, is
the question of the physical interpretation of the conditions (\ref{decouplCond1}).
It is clear, for example, that if the distortions to the initially uncoupled
beam matrix $\Sigma$ were produced by a magnetostatic system, then the decoupling
also can be done by a magnetostatic system, but how such beam matrices can be 
described more intuitively and what are the other possibilities?
Currently, as more physical example in the comparison with the conditions 
(\ref{decouplCond1}) description, we only can state that all beam matrices with 
equal eigenemittances (definition and properties of eigenemittances can be found 
in \cite{MomInv_4,MyEE}) always can be decoupled by a magnetostatic system. 
It follows from the observation that the conditions (\ref{decouplCond1}) are equivalent 
to the property of the matrix $(\Sigma J_6)^2$ to have zeros in the positions

\noindent
\begin{eqnarray}
(\Sigma J_6)^2 = 
\left(
\begin{array}{cccccc}
* & * & * & * & 0 & *\\
* & * & * & * & 0 & *\\
* & * & * & * & 0 & *\\
* & * & * & * & 0 & *\\
* & * & * & * & * & *\\
0 & 0 & 0 & 0 & * & *
\end{array}
\right),
\label{CTTLD01}
\end{eqnarray}

\noindent
and from the fact proven in \cite{My02} that if the matrix $\Sigma$ has
all eigenemittances equal to each other, then the matrix $(\Sigma J_6)^2$
is a diagonal matrix.

\subsection{Illustrative example}

We have seen that if in the beam matrix $\Sigma$ there are nonzero correlations 
between energy of particles and their transverse positions and momenta, then the 
values of the transverse projected emittances can be reduced, but how these reduced 
emittances are related to the emittances of the particle beam before it was damaged 
by the CSR wake (or by some other effects) remains, of course, completely unclear.
So, let us consider an example which would give at least some insights into this problem.

Let us assume that we have in the beginning a particle beam with the beam matrix 
$\Sigma$ in which all degrees of freedom are decoupled from each other

\noindent
\begin{eqnarray}
\Sigma = 
\left(
\begin{array}{cccccc}
\langle x^2                \rangle & \langle x p_x              \rangle & 0 & 0 & 0 & 0\\
\langle x p_x              \rangle & \langle p_x^2              \rangle & 0 & 0 & 0 & 0\\
0 & 0 & \langle y^2                \rangle & \langle y p_y              \rangle & 0 & 0\\
0 & 0 & \langle y p_y              \rangle & \langle p_y^2              \rangle & 0 & 0\\
0 & 0 & 0 & 0 & \langle \sigma^2           \rangle & \langle \sigma \varepsilon \rangle\\
0 & 0 & 0 & 0 & \langle \sigma \varepsilon \rangle & \langle \varepsilon^2      \rangle
\end{array}
\right),
\label{Form1960}
\end{eqnarray}

\noindent
and then this beam passes through a beamline described by the matrix 

\noindent
\begin{eqnarray}
T = 
\left(
\begin{array}{cccccc}
1 & 0 & 0 & 0 & 0 & 0\\
0 & 1 & 0 & 0 & a & 0\\
0 & 0 & 1 & 0 & 0 & 0\\
0 & 0 & 0 & 1 & 0 & 0\\
0 & 0 & 0 & 0 & 1 & 0\\
a & 0 & 0 & 0 & b & 1
\end{array}
\right),
\quad
a \neq 0.
\label{Example01}
\end{eqnarray}

\noindent
Our choice of the matrix $T$ as a source of the growth of the projected emittances 
and also as a source of the transverse to longitudinal coupling is motivated by the 
following reasons: The matrix $T$, from one side, is symplectic and therefore all changes 
which it introduces are reversible, but it is not the matrix of a magnetostatic system
and it is interesting to see up to what extend the original projected emittances 
of the matrix $(\ref{Form1960})$ can be recovered afterwards by a magnetostatic 
correction system. From the other side, this matrix, similar to the wake field 
action, provides transverse kick and energy loss to the particle depending on 
its longitudinal position within the bunch. Note that if parameters $a$ and $b$ 
in this matrix are related to each other in some special way, then the matrix $T$ 
becomes equal to the matrix of the thick-lens horizontally deflecting cavity when it
is sandwiched between two drifts of equal negative lengths (see, for example, \cite{CornEmma}).

The passage of the beam matrix (\ref{Form1960}) through the system described by the 
matrix $T$ gives equal increase of horizontal and longitudinal projected emittances 
(the vertical degree of freedom remains decoupled from the others and is ignored in
the following considerations)

\noindent
\begin{eqnarray}
\varepsilon_x^2 \, \leftarrow \,
\varepsilon_x^2 \, + \, a^2 \, \langle x^2  \rangle \,
\langle \sigma^2 \rangle,
\label{Example02}
\end{eqnarray}

\noindent
\begin{eqnarray}
\varepsilon_{\sigma}^2 \, \leftarrow \,
\varepsilon_{\sigma}^2 \, + \, a^2 \, \langle x^2  \rangle \,
\langle \sigma^2 \rangle,
\label{Example03}
\end{eqnarray}

\noindent
and generates horizontal to longitudinal coupling 
terms (beam dispersions and beam tilts)

\noindent
\begin{eqnarray}
\langle x \varepsilon  \rangle \, \leftarrow \, a \, \langle x^2 \rangle,
\label{Example04}
\end{eqnarray}

\noindent
\begin{eqnarray}
\langle p_x \varepsilon  \rangle \, \leftarrow \, 
a 
\,
\left(
\langle x p_x \rangle \, + \, \langle \sigma \varepsilon \rangle
\, + \, b \, \langle \sigma^2 \rangle
\right),
\label{Example05}
\end{eqnarray}

\noindent
\begin{eqnarray}
\langle x \sigma  \rangle \, \leftarrow \, 0,
\label{Example06}
\end{eqnarray}

\noindent
\begin{eqnarray}
\langle p_x \sigma  \rangle \, \leftarrow \, a \, \langle \sigma^2 \rangle.
\label{Example07}
\end{eqnarray}

\noindent
The rms bunch length squared $\langle \sigma^2 \rangle$ is conserved, 
but the rms energy spread evolves according to the formula

\noindent
\begin{eqnarray}
\langle \varepsilon^2 \rangle \, \leftarrow \, \mbox{\ae}, 
\label{Example08}
\end{eqnarray}

\noindent
where

\noindent
\begin{eqnarray}
\mbox{\ae} = \langle \varepsilon^2 \rangle \, + \, a^2 \, \langle x^2 \rangle
\, + \, 2 \, b \, \langle \sigma \varepsilon \rangle
\, + \, b^2 \, \langle \sigma^2 \rangle \, > \, 0,
\label{Example088}
\end{eqnarray}

\noindent
and the beam energy chirp also experiences some change

\noindent
\begin{eqnarray}
\langle \sigma \varepsilon \rangle \, \leftarrow \, 
\langle \sigma \varepsilon  \rangle \, + \, b \, \langle \sigma^2 \rangle.
\label{Example09}
\end{eqnarray}

The equations (\ref{decouplCond1}), when applied to the matrix
$T \Sigma T^{\top}$, are reduced to the single relation

\noindent
\begin{eqnarray}
\varepsilon_x \,=\, \varepsilon_{\sigma}
\label{Example099}
\end{eqnarray}

\noindent
among the elements of the matrix $\Sigma$. It means that both projected emittances can be 
recovered by a magnetostatic correction (and also the beam matrix can be decoupled) if and 
only if horizontal and longitudinal projected emittances were equal in the beginning before 
the passage through the system described by the matrix $T$. But let us see what can be done 
if they were not. So, as the next step, let the beam pass through the dispersive part of the 
downstream correction system and, because we would like to express the final results using 
the elements of the original matrix (\ref{Form1960}), let us consider the transformation

\noindent
\begin{eqnarray}
\Sigma \leftarrow
(R_2 T) \, \Sigma \, (R_2 T)^{\top},
\label{Exam009}
\end{eqnarray}

\noindent
instead of the transformation (\ref{S01}).

The formulas (\ref{evolutionEX}) and (\ref{evolutionEZ}) for the transport of 
the horizontal and longitudinal projected emittances, when adapted to the 
transport equation (\ref{Exam009}), can be rewritten as follows:

\noindent
\begin{eqnarray}
\varepsilon_x^2 
\,\leftarrow\, 
\varepsilon_x^2
\,+\,
\Psi_F(r_{51}^x - r_{51}, \, r_{52}^x - r_{52})
\,-\,
\Psi_F^x,
\label{tmp01}
\end{eqnarray}

\noindent
where

\noindent
\begin{eqnarray}
r_{51}^x \,=\, a \,
\frac{
\langle x p_x \rangle + \langle \sigma \varepsilon \rangle + 
b \, \langle \sigma^2 \rangle}{\mbox{\ae}},
\label{tmp02}
\end{eqnarray}

\noindent
\begin{eqnarray}
r_{52}^x \,=\, - a \, \frac{\langle x^2 \rangle}{\mbox{\ae}},
\label{tmp03}
\end{eqnarray}

\noindent
and

\noindent
\begin{eqnarray}
\Psi_F^x \,=\,
-
\frac{a^2 \langle x^2  \rangle}{\mbox{\ae}} \,
\left( \varepsilon_{\sigma}^2 \,-\, \varepsilon_x^2 \right).
\label{tmp04}
\end{eqnarray}

\noindent
\begin{eqnarray}
\varepsilon_{\sigma}^2 
\,\leftarrow\, 
\varepsilon_{\sigma}^2
\,+\,
\Psi_F(r_{51}^{\sigma} - r_{51}, \, r_{52}^{\sigma} - r_{52})
\,-\,
\Psi_F^{\sigma},
\label{tmp05}
\end{eqnarray}

\noindent
where

\noindent
\begin{eqnarray}
r_{51}^{\sigma} \,=\, a \,
\frac{
\left(\langle \sigma \varepsilon \rangle  + b \, \langle \sigma^2 \rangle \right) 
\, \varepsilon_x^2 \, + \,
\langle x p_x \rangle \, \varepsilon_{\sigma}^2}{\varkappa},
\label{tmp06}
\end{eqnarray}

\noindent
\begin{eqnarray}
r_{52}^{\sigma} = -a \,
\frac{
\langle x^2 \rangle \, \varepsilon_{\sigma}^2}{\varkappa},
\label{tmp07}
\end{eqnarray}

\noindent
\begin{eqnarray}
\Psi_F^{\sigma} \,=\,
\frac{a^2 \langle x^2 \rangle \, \varepsilon_{\sigma}^2}{\varkappa} \,
\left( \varepsilon_{\sigma}^2  \,-\, \varepsilon_x^2 \right),
\label{tmp08}
\end{eqnarray}

\noindent
and

\noindent
\begin{eqnarray}
\varkappa \,=\,
a^2 \langle x^2 \rangle \left( \varepsilon_{\sigma}^2 \,-\, \varepsilon_x^2 \right)
+ \mbox{\ae} \, \varepsilon_x^2 \,>\, 0.
\label{tmp09}
\end{eqnarray}

\noindent
Note that in the above formulas $\Psi_F$ is a positive definite quadratic form
in two variables obtained from the quadratic form $\Psi_A$, and the exact
expression for the $2 \times 2$ matrix associated with the quadratic form 
$\Psi_F$ is unimportant for the further consideration.

From the equation (\ref{tmp01}) one sees that the original horizontal
projected emittance $\varepsilon_x$ can be recovered if and only if

\noindent
\begin{eqnarray}
\Psi_F^x \,\geq\, 0
\quad
\Leftrightarrow
\quad
\varepsilon_x \,\geq\, \varepsilon_{\sigma},
\label{geom01}
\end{eqnarray}

\noindent
and the condition for the recovering of $\varepsilon_{\sigma}$ coming
from the equation (\ref{tmp05}) is

\noindent
\begin{eqnarray}
\Psi_F^{\sigma} \,\geq\, 0
\quad
\Leftrightarrow
\quad
\varepsilon_{\sigma} \,\geq\, \varepsilon_x.
\label{geom02}
\end{eqnarray}

\noindent
So, as one sees, if $\varepsilon_x \neq \varepsilon_{\sigma}$, then only the larger 
of the two can be repaired and even can be further reduced, but only on expense of 
the increase of the other. Nevertheless, even if the horizontal (or longitudinal) 
projected emittance cannot be recovered to its original value, the distorted value
(\ref{Example02}) (or (\ref{Example03})) always can be reduced, as follows 
from the theory developed in this paper.

Let us now consider three extreme cases: solution for the lattice dispersions
which minimizes $\varepsilon_x$, solution which minimizes $\varepsilon_{\sigma}$
and solution which zeros beam tilts. Even before making any calculations, one can
state that in all these three cases the sum

\noindent
\begin{eqnarray}
\varepsilon_x^2 \,+\, \varepsilon_{\sigma}^2
\label{geom03}
\end{eqnarray}

\noindent
will be conserved, which follows from the preservation of the Lysenko 
invariant (\ref{LysenkoInvariant}) and the fact that all these solutions
make the vector of the beam dispersions and the vector of the beam tilts
linearly dependent at the correction system exit. 
Note also that due to the conservation of the sum (\ref{geom03})
and due to the extremum properties of the solutions which minimize projected
emittances, any solution which zeros beam tilts will give the final value
of the transverse projected emittance which must lie 
between the values given by the solution which zeros beam dispersions and 
the solution which minimizes longitudinal projected emittance.

The setting of the lattice dispersions to the values $r_{51}=r_{51}^x$ and
$r_{52}=r_{52}^x$ which minimize the horizontal projected emittance $\varepsilon_x$ 
gives us

\noindent
\begin{eqnarray}
\varepsilon_x^2 \, \leftarrow \, \varepsilon_x^2
\, + \, \frac{a^2 \langle x^2  \rangle}{\mbox{\ae}} \,
\left( \varepsilon_{\sigma}^2 \, - \, \varepsilon_x^2 \right),
\label{Example11}
\end{eqnarray}

\noindent
\begin{eqnarray}
\varepsilon_{\sigma}^2 \, \leftarrow \, \varepsilon_{\sigma}^2
\, - \, \frac{a^2 \langle x^2  \rangle}{\mbox{\ae}} \,
\left(\varepsilon_{\sigma}^2 \,-\, \varepsilon_x^2\right),
\label{Example12}
\end{eqnarray}

\noindent
and minimization of the longitudinal projected emittance $\varepsilon_{\sigma}$ 
by the setting $r_{51}=r_{51}^{\sigma}$ and $r_{52}=r_{52}^{\sigma}$ produces 

\noindent
\begin{eqnarray}
\varepsilon_x^2 \, \leftarrow \, \varepsilon_x^2
\, + \, \frac{a^2 \langle x^2 \rangle \, \varepsilon_{\sigma}^2}{\varkappa} \,
\left( \varepsilon_{\sigma}^2 \, - \, \varepsilon_x^2 \right),
\label{lonem02}
\end{eqnarray}

\noindent
\begin{eqnarray}
\varepsilon_{\sigma}^2 \, \leftarrow \, \varepsilon_{\sigma}^2
\, - \, \frac{a^2 \langle x^2 \rangle \, \varepsilon_{\sigma}^2}{\varkappa} \,
\left(\varepsilon_{\sigma}^2 \,-\, \varepsilon_x^2\right).
\label{lonem03}
\end{eqnarray}

\noindent
Unfortunately, it is practically impossible to find the general solutions
for the lattice dispersions which are required for the zeroing of the beam 
tilts in the analytical form, and we will give it only for the partial
case when $r_{56} = b = 0$. With this assumption the solution for the
zeroing of the beam tilts is unique and is given by the following formulas

\noindent
\begin{eqnarray}
r_{51} \, = \,
\frac{a 
\langle \sigma^2 \rangle 
\left(
\langle x p_x  \rangle 
+
\langle \sigma \varepsilon \rangle 
\right)}{\kappa},
\label{ExR4}
\end{eqnarray}

\noindent
\begin{eqnarray}
r_{52} \, = \,
-
\frac{a 
\langle x^2 \rangle
\langle \sigma^2 \rangle 
}
{\kappa},
\label{ExR402}
\end{eqnarray}

\noindent
where

\noindent
\begin{eqnarray}
\kappa \, = \,
\varepsilon_x^2 
\, + \,
a^2 \langle x^2 \rangle
\langle \sigma^2 \rangle
\, + \,
\langle \sigma \varepsilon  \rangle^2,
\label{ExR5}
\end{eqnarray}

\noindent
and the resulting formulas for the transport of the projected emittances are

\noindent
\begin{eqnarray}
\varepsilon_x^2 
\,\leftarrow\,
\varepsilon_x^2
\,+\, 
\frac{a^2 \langle x^2  \rangle \langle \sigma^2  \rangle}{\kappa}
\,
\left(\varepsilon_{\sigma}^2 \,-\, \varepsilon_x^2 \right),
\label{ExR8}
\end{eqnarray}

\noindent
\begin{eqnarray}
\varepsilon_{\sigma}^2 
\,\leftarrow\,
\varepsilon_{\sigma}^2
\,-\, 
\frac{a^2 \langle x^2  \rangle \langle \sigma^2  \rangle}{\kappa}
\,
\left(\varepsilon_{\sigma}^2 \,-\, \varepsilon_x^2\right).
\label{ExR9}
\end{eqnarray}

\noindent
One can check that while for $a \neq 0$ the result of (\ref{Example11})
is always smaller than the result of (\ref{Example02}) (as expected), 
the values produced by the formulas (\ref{lonem02}) and (\ref{ExR8}) for 
$\varepsilon_x \neq \varepsilon_{\sigma}$ can reduce the distorted emittance 
(\ref{Example02}) only under specific conditions.

\begin{acknowledgments}
The authors are thankful to Marc Guetg for valuable discussions.
\end{acknowledgments}

\end{document}